# Automatically Annotating Articles Towards Opening and Reusing Transparent Peer Reviews

**Afshin Sadeghi** [1,2]*****, **Sarven Capadisli** [1] **and Johannes Wilm** [3,], **Christoph Lange**[1,2], **Philipp Mayr**[3]

[1] University of Bonn, Germany; sadeghi@cs.uni-bonn.com , info@csarven.ca, langec@cs.uni-bonn.de
[2] Fraunhofer IAIS, Germany
[3] GESIS – Leibniz Institute for the Social Sciences, Germany; mail@johanneswilm.org, philipp.mayr@gesis.org
***** Correspondence: sadeghi@cs.uni-bonn.com

**Abstract:** An increasing number of scientific publications are created in open and transparent peer review models: a submission is published first, and then reviewers are invited, or a submission is reviewed in a closed environment but then these reviews are published with the final article, or combinations of these. Reasons for open peer review include giving better credit to reviewers, and enabling readers to better appraise the quality of a publication. In most cases, the full, unstructured text of an open review is published next to the full, unstructured text of the article reviewed. This approach prevents human readers from getting a quick impression of the quality of *parts* of an article, and it does not easily support secondary exploitation, e.g., for scientometrics on reviews. While document formats have been proposed for publishing structured articles including reviews, integrated tool support for entire open peer review workflows resulting in such documents is still scarce. We present **AR-Annotator**, the Automatic **Article and Review Annotator** which employs a semantic information model of an article and its reviews, using semantic markup and unique identifiers for all entities of interest. The fine-grained article structure is not only exposed to authors and reviewers but also preserved in the published version. We publish articles and their reviews in a Linked Data representation and thus maximise their reusability by third party applications. We demonstrate this reusability by running quality-related queries against the structured representation of articles and their reviews.

## 0. Introduction

The scientific community expects that the quality of published articles is controlled in a review process. When an author has finished writing an article, it is typically reviewed by two or more independent experts. Successfully passing the review process is a prerequisite for an article to be published. Besides generally recommending whether or not the article in its current shape has a sufficient quality to be published, reviewers give specific feedback on parts of the article, often including detailed comments suggesting improvements. When authors are notified of reviews, they take them into account in their next revision of the article. Traditionally, reviews have mostly been confidential, whereas today there is a strong movement towards *open* peer reviews. We discuss the benefits of open peer reviews in subsection 0.1, state the lack of tool support for publishing open reviews as a problem in subsection 0.2, and introduce the structure of the rest of this article in subsection 0.3.

*0.1. Open Peer Review*

In traditional workflows, reviews are confidential. The programme committee chair of a conference or the editor of a journal initially receives the reviews and passes them on to the authors. The identity of the reviewers is typically not disclosed to the authors ("single-blind review"); in some communities, the reviewers do not know the identities of the authors either ("double-blind review"). Nowadays there is a strong movement towards *open peer reviews* in many communities [1]. One common variant starts with a traditional closed process, after whose end the reviews (including the identities of the reviewers) are published (sometimes with the possibility for reviewers to opt out); in another variant, the unreviewed article is published first and then reviewed. The open peer review policy employed by the *Semantic Web Journal*, for example, aims to "minimize positive and negative bias in the reviewing" [2]. *BMJ Open*[1] is another journal that publishes all the reviews alongside the original article, as well as the authors' responses to the reviews. Another pioneer in open scholarly publishing is Data Science Journal[2], which not only discloses the reviews of articles but also publishes the responses of authors to the comments and decisions of the reviewers. Some studies suggest a positive effect of this openness on the quality of reviews [3,4]. Further reasons for open peer review include giving better credit to the work done by the reviewers, as well as enabling readers to better appraise the quality of a publication.

*0.2. Problem Statement: Lack of Tool Support*

Review processes are technically well supported by submission and review management systems such as EasyChair[3] for conferences, or Open Journal Systems[4] for journals. However, these systems focus on facilitating the assignment of reviewers to submissions and on automating the sending of notifications by email. Besides requesting the client to enter some global numeric scores, they treat the actual review as an unstructured block of text. When such a review is entered into the system, any explicit information on the connection of reviewers' comments to parts of the article is lost. Both the authors, aiming at understanding where they should improve their article, and readers of such a review published openly have to read the review in full and switch back and forth between the review and the article in order to understand the connection. Even if reviewers point to precise text locations, such as "the caption of Figure 1" or "the last sentence on page 7" (which is tedious without software support), it takes readers (including the authors as a special case) time, and it is error-prone to interpret these references manually. State-of-the-art tools for reading and writing articles support comments anchored to precise ranges of text, thus eliminating any ambiguity regarding what part of the article a reviewer's comment refers to. Submission and review management systems are often configured to accept the upload of attachments, which could be a version of the article annotated with such comments. If such facilities were used more frequently, such reviews published openly would be easier to consume for human readers than traditional ones – as discussed in section 4, but their utility would still remain limited from the perspective of *automated* approaches for *secondary exploitation* of reviews, e.g., for giving credit to reviewers or for appraising the quality of an article. For example, it would still require human intelligence to understand whether the authors' observations from the evaluation of a system were accepted by the reviewers or discussed controversially. Document formats such as HTML+RDFa[5], combined with suitable vocabularies such as the SPAR (Semantic Publishing and Referencing) ontologies[6] for making the structure of the article explicit, or the Web Annotation Vocabulary[7] for making the provenance of comments and their connection to the text explicit, support the publication of articles and reviews in a way that appeals both to human readers and to machine services. However, document editors as well as submission and review management systems addressing end users without a web and knowledge engineering background have so far not supported such formats: neither have document editors supported an explicit representation of the structure of an article (such as "this is the evaluation section"), nor have review management systems supported human as well as machine friendly publication of reviewers' comments attached to parts of articles in a fine-grained way.

The contributions of this work are as follows:

- A pipeline, named AR-Annotator (Article and Review Annotator), to enrich articles and reviews by with RDFa data, immediately preparing them for efficient analysis by SPARQL queries, and furthermore

---

[1] http://bmjopen.bmj.com/pages/reviewerguidelines/
[2] https://datasciencehub.net
[3] http://www.easychair.org
[4] https://pkp.sfu.ca/ojs/
[5] http://www.w3.org/TR/rdfa-syntax
[6] http://www.sparontologies.net
[7] https://www.w3.org/TR/annotation-vocab/

- providing a way to allow them to be added to the LOD Cloud and scholarly communication knowledge graphs such as SCM-KG [5].
- We support the repeatability and reproducibility of our approach and the empirical evaluation results by providing the source code of the article enriching module and the evaluation procedure under Apache license.[8]

*0.3. Structure of this Article*

The remainder of this article is structured as follows: section 1 presents concrete use cases for a system that publishes peer-reviewed articles on the Web appealing to human and machine clients, and derives precise requirements from these use cases. section 2 discusses related approaches specifically from the perspective of how well they support our use cases and how well they address our derived requirements. section 3 presents our implementation, which supports the given use cases and addresses the given requirements. It comprises a pipeline in which Fidus Writer, Open Journal Systems (OJS), and dokieli are connected by the core of AR-Annotator, which translates a commented document resulting from the combined usage of Fidus Writer and OJS to dokieli's HTML+RDFa article format, preserving the original structure of the document but further enriching it with extracted semantics. section 4 Demonstrates the usefulness of attaching reviews to documents in a study.

section 5 concludes with an outlook to future work and a sketch of the mid-term impact that a wide adoption of AR-Annotator could make on scholarly communication.

**1. Use Cases and Derived Requirements**

subsection 1.1 presents concrete use cases for a system that publishes peer-reviewed articles on the Web appealing to human and machine clients. subsection 1.2 derives precise requirements from these use cases. section 4 presents a plausibility test that justifies the specific requirement to attach reviewers' comments to parts of articles in a fine-grained way by proving the usefulness of doing so.

*1.1. Use Cases*

We first outline use cases our approach aims to support. Overall, open science aims at supporting the reuse of research output and making transparent the advancement of human knowledge using the Web. We aim at pushing this approach by better utilizing the interactivity potential of Web technology to realize an integrated authoring, reviewing and publishing workflow.

Soliciting Additional Reviews:

We aim at publishing articles together with their reviews, in the first step focusing on reviews coming from a closed workflow. Nevertheless, we would like the published article to attract further comments from new reviewers in the community, even after acceptance. This is helpful in settings when a sufficient number of reviews cannot be obtained by traditional means, or, more generally, when the authors but also readers welcome additional feedback.

Automated Analysis of Reviews:

Several scenarios could benefit from an automated analysis of reviews by repeatable queries against the review comments and the structure of the article. Not only would this help the primary recipients of reviews, i.e., the authors, to faster comprehend the reviewers' feedback when revising their article, by answering queries such as "what section does have most comments attached to?". It would also help editors of a journal, or chairs or senior PC members of a conference, to realize commonalities and differences between the comments of the original reviewers, e.g., "what statements were commented on by all reviewers?". Finally, such queries could give rise to new scientometrics based on secondary exploitation of reviews, e.g., "to what extent do reviewers disagree with the 'Methods' section?".

---

[8] https://github.com/OSCOSS/AR-Annotator/

Sharing and Reusing Articles and Reviews:

To support further reuse of articles and reviews, e.g., by sharing them in social networks or by enabling precise citations, we would like each of their structural components (sections, figures, references, comments, etc.), including each reviewer's comment, to be identifiable – and thus linkable – in a globally unique way and to carry its own local metadata.

Multi-device Accessibility of Reviews:

By publishing reviews using Web technology, we aim at increasing their accessibility by making them readable across devices, including devices with different screen sizes, as well as different output modalities including text and speech, and by making their appeAR-Annotatorce, e.g., font size, adaptable to the requirements of the reader.

Independent Quality (Re-)Assessment:

Making explicit the structure of an article can support the review process by giving reviewers easier access to the resources underlying an article, such as research datasets and software artifacts, thus facilitating the reproducibility of research and reducing the effort of an independent (re-)assessment of an article's quality.

*1.2. Requirements*

From the use cases outlined so far, we derived the following technical requirements for a system supporting the workflow of authoring, reviewing and publishing (in this order):

- label=R0 It should be possible to create initial reviews for an article in a closed submission and review management system.
- lbbel=R0 The support for comments and annotations is the most requested feature from the reviewing tools [6]. Reviewers should be enabled to attach comments in a fine-grained way to precise structural elements of an article, and this attachment should be preserved in the published document.
- lcbel=R0 The structure of all parts of an article should be exposed explicitly; each structural component of an article should have a globally unique identifier.
- ldbel=R0 Not only should the structure of the article from the authoring environment be preserved, but, where possible, structural information implicitly hidden in unstructured text should be extracted, and the text of the published article should be enriched with an explicit representation of such structures.
- lebel=R0 Our system should publish reviews in a standard web format, which humans can consume with a browser, but which can also carry semantic metadata enabling queries and other automated analyses.
- lfbel=R0 It should be possible to add further reviews to the published document.

Besides legal openness in terms of open access, these requirements enable openness in a technical sense.

As an additional requirement, we aim at compatibility with traditional workflows. We do not aim at entirely disrupting workflows. It should remain possible to subsequently open up articles and reviews created in a traditional, closed workflow.

## 2. Related Work

We first review related work focusing on tool support for open peer review, then studies that have targeted the annotation of articles.

*2.1. Tool Support for Open Peer Review*

With an increasing adoption of open peer review, as explained in subsection 0.1, the situation has improved over 2012, when Kriegeskorte argued that the leading review systems hardly published general comments about articles [7]. However, CL@AS: This statement might be criticized, as we didn't systematically review a significant number of open peer review journals. However, doing so would be expensive, and my intuition is that it wouldn't yield a different result. Therefore, let's leave this statement as is. AS@CL I totally understand and agree with the concern.none of the open peer review journals we have reviewed provide technical support for publishing reviews with fine-grained comments attached to parts of articles. To the best of our knowledge, state-of-the-art online open publishing systems do not offer advanced facilities for publishing articles and reviews in a way that makes them easy to consume for humans, let alone machines. For example, the client-server *Open Journal Systems*[9] (OJS), supports a sophisticated reviewing workflow management featuring different user roles and access permissions as well as the publication of reviews once they have been finished, but only as a block of text in the static PDF format.

To enable open peer review with fine-grained comments in a "publish first" setting, one could simply employ collaborative web-based authoring tools with commenting facilities, such as Google Docs[10], which targets non-technical users. However, this approach does not support export of reviews and it is not integrated with any review workflow. In our previous work, we have integrated Fidus Writer[11], an academically oriented alternative to Google Docs, with OJS to combine collaborative authoring and a sophisticated management of a traditional peer review workflow [8] – but this combination does not support the publication of articles annotated with fine-grained reviewers' comments because the articles in a Fidus Writer installation can only be viewed by other users registered there, while online publishing is only possible by exporting, e.g., to HTML, which loses the comments.

*dokieli*[12], a web-based tool for editing and publishing articles in human as well as machine-friendly semantic HTML+RDFa markup developed by the second author, supports commenting published articles in a decentral architecture [9]. However, it does not provide dedicated support for the scholarly reviewing workflow *before* publishing.

*2.2. Tools for Annotating Articles*

dokieli enables linking to structural parts of articles and to comments, and it supports *manual* enrichment of articles with semantic annotations. Other Web-based tools that support authors with annotation facilities but not specifically tailored for scientific articles include Loomp [10], a general-purpose annotation framework targeted at journalists, which does not perform automatic annotation either, and the RDFaCE [11] RDFa content editor, which supports automated annotation of news articles.

Several systems support automatic annotation of the content of articles from specific fields of science with ontologies for the respective domains. For example, DOMEO [12] and BioC [13] use text mining services to analyze the bodies of articles and to relate them to biomedical ontologies. The more recent AnnoSys supports generic annotations of articles [14]. However, it supports Web publishing and annotates in a local, fine-grained way, which would facilitate consumption, it publishes the annotations as separate XML documents in contrast to dokieli, which embeds them as RDFa. The ACM Article Content Miner [15] uses the SPAR ontologies to extract classes that describe the structure of articles. Although mining for SPAR-related keywords allows extracting author affiliations, references, etc. from articles, ACM keeps the extracted information separated from the articles.

## 3. Implementation

This section presents AR-Annotator (Article and Review Annotator), the implementation of our proposed approach, beginning with the overall architecture of a pipeline comprising the steps of authoring, reviewing and publishing in subsection 3.1, then, in subsection 3.2, a description of the previously existing components of which our implementation of the pipeline is composed – the Fidus Writer document editor for authoring and commenting, Open Journal Systems (OJS) for managing the review process, and dokieli for publishing the reviewed article on the Web –, and finally, in subsection 3.3, the core of AR-Annotator, which translates a commented document resulting from the combined usage of Fidus Writer and OJS to dokieli's HTML+RDFa format, preserving its original structure and further enriching it by an explicit representation of additional structural aspects.

---

[9] https://pkp.sfu.ca/ojs/
[10] https://docs.google.com/
[11] https://www.fiduswriter.org/
[12] https://dokie.li/

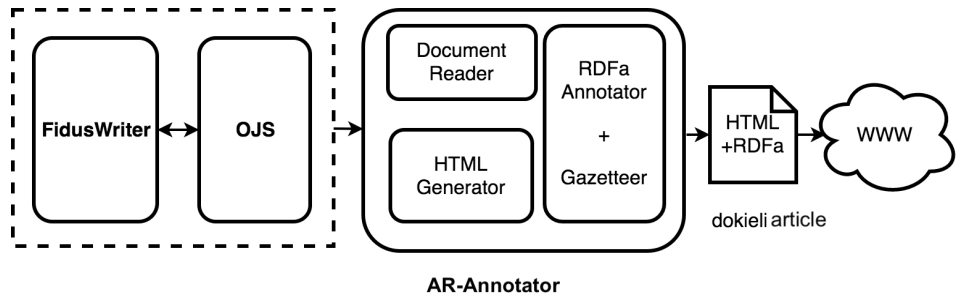

**Figure 1.** A pipeline for scholarly article authoring, reviewing and publishing facilitated by AR-Annotator

*3.1. Architecture*

Figure 1 provides a conceptual overview of the pipeline supported by AR-Annotator for authoring, reviewing and publishing scholarly articles. The AR-Annotator module is located between a tool for writing articles and the Web as the final publication infrastructure. It receives an article as a document with formatted text, tables, images and reviews belonging to it and exports a semantically enriched article in semantic Web-compatible markup, which is ready for human and machine consumption.

*3.2. Integration of Reused Components: Fidus Writer, OJS and dokieli*

This subsection presents the existing components that complete the AR-Annotator pipeline and explains why we chose them. We argue how transforming an article written in Fidus Writer to dokieli's HTML+RDFa format enables open scientific publishing and reviewing. Going beyond the discussion of the previous state of these tools in section 2, we discuss how the AR-Annotator core fills the gap between them by implementing this transformation step.

The initial, closed review is conducted via a centrally hosted combination of Fidus Writer and Open Journal Systems (OJS) [8], which allows authors, reviewers and journal editors to cooperate seamlessly across the boundary of the two systems after only registering once. This integration frees the authors from having to export their article from Fidus Writer, e.g., to PDF, and to upload that copy to OJS, and from once more having to download the article once it has been reviewed. Instead, the reviewers work in the original collaborative Fidus Writer document. After the end of the review process, their comments become visible to the authors right in place.

As discussed in subsection 2.1, the Fidus Writer/OJS combination does not support high-quality online publication of articles that contain reviews. dokieli documents, on the other hand, can be displayed in a visually appealing way thanks to supporting the styles of publishers (currently Springer and ACM). They may include comments as well as machine-comprehensible RDFa metadata, making explicit the structure of an article. Furthermore, when users have access to appropriately configured servers, which may even be provided decentrally, they can add further public comments to a dokieli article. However, dokieli alone does not support the management of a review process in a way as sophisticated as OJS. Thus, to benefit from the best features of both systems, we combine them by extending Fidus Writer, which has so far only featured a simple HTML export filter as well as LaTeX, ODT and DOCX export filters, all without support for comments, with a facility that exports to the HTML+RDFa format of dokieli, as detailed below in subsection 3.3.

The conservative Fidus Writer/OJS approach risks moving publishing to the Web without taking advantage of the opportunity to reform parts of the publishing process that no longer correspond to the current state of the development of the means of communication, ultimately limiting the review process to a small number of academics with privileged access to a central infrastructure. dokieli articles, on the other hand, can have their own decentral "inboxes", i.e., containers in which they can receive notifications, e.g., from reviewers' comments or other annotation activities.

| Criterion | Fidus Writer | dokieli |
|---|---|---|
| Discoverability of reviews | N/A | Yes (RDFa) |
| Machine-comprehensible representation | N/A | Yes |
| Creation and access to reviews | post publication on article snapshot | pre or post publication (optionally on snapshots) |
| Systematic traditional reviewing workflow | Yes, by OJS integration closed phase reviews is supported | N/A |
| Multiple reviewing rounds | Yes, by OJS integration | NA |
| User-centered publishing possibility | Restricted to PDF | Web based adaptive rendering |
| Publishing reviews with articles | N/A | Yes, over the Web |

**Table 1.** Strengths of the selected components w.r.t. our requirements

Table 1 highlights the respective strengths of the two selected components with regard to criteria corresponding to the requirements that we have derived from our use cases in section 1. It is obvious that the strengths are complementary and that therefore the integration of both components fully supports the desired workflow.

*3.3. The AR-Annotator Article and Review Annotator*

As a translation between Fidus Writer/OJS and dokieli, we implemented AR-Annotator. Figure 2 shows a sample article including reviewers' comments in Fidus Writer on top, and its export to dokieli below while a user is adding a second review comment using dokieli's commenting functionality. The dokieli HTML+RDFa is generated in two high-level steps:

1. Transformation of the article's explicit structure, including metadata and comments, to dokieli's HTML+RDFa format.
2. Identification of structure (discourse elements) in the unstructured text of the article, and enrichment of the article by an explicit representation of this structure.

For example, step 1 gives us the basic structure of a section in the article, whereas step 2 marks the *subject* of that section, e.g., a section titled "Introduction" will be annotated as an instance of *deo:Introduction*, a class from the "Discourse Elements Ontology" of the SPAR family.

Three modules in AR-Annotator realize this workflow. First, a **Document Reader** reads document elements of Fidus Writer articles individually. To stay faithful to the content of the original article, it records all the article elements and their metadata. This metadata includes the list of authors, table and image features, the types of textual elements, and comments. To preserve the structure of the original article it reads the article sequentially from top to bottom, keeping the order of appeAR-Annotatorce of elements.

The second module is the **HTML Generator**. To realize this part, we analyzed the structure of articles written in dokieli and matched the semantics of each document element read from Fidus Writer to their target presentation in dokieli. To gain an equivalent HTML structure of the input article, this module once more produces HTML elements sequentially. It defines proper dokieli HTML patterns based on the metadata provided by document reader. Into these HTML patterns, we include a representation of the structural information as RDFa, representing, e.g., comments as Web Annotations.

The third module, the **RDFa Annotator**, works closely with the HTML generator of step 2. This module performs named entity recognition on common information units – including article title, abstract, section headers, author names, references, images and tables – in the article and then maps them to corresponding RDFa patterns using suitable RDF vocabularies. To create this list of keywords, called gazetteer in natural language processing, we reviewed section titles of 40 articles[13] in the field of computer science that used a variety of terminologies and article structures, took their title sections, and mapped them to entities in our ontology. This set comprises names of entities such as "Introduction and Motivation", "Conclusion", etc. We mapped these keywords to classes defined in the *schema.org*[14], *DEO* (Discourse Elements Ontology)[15] and *SWRC*[16] ontologies.

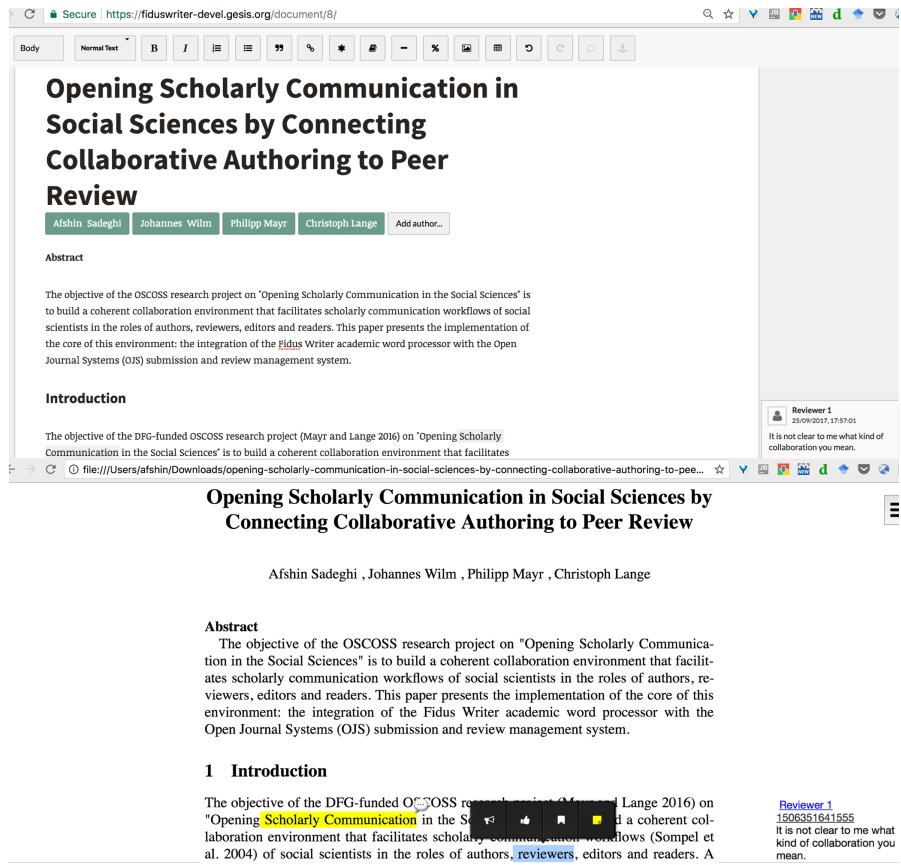

**Figure 2.** A reviewed article in Fidus Writer (top) and dokieli (bottom)

AR-Annotator finally creates a self-contained ZIP archive containing the HTML+RDFa article, separate HTML files (embedded into the article) including the review comments, as well as accompanying media files. This archive is suitable for deploying on any server and also for offline use.

For evaluation, we provide a sample article created by our approach[17]. We also implemented a plugin to invoke our approach from Fidus Writer.[18] Although users can edit the automatically generated annotations in the resulting dokieli document, this plugin also supports a user-friendlier manual assignment of annotations in Fidus Writer before running the export.

**4. Usefulness of publishing fine-grained review comments**

The requirement of supporting fine-grained reviewers' comments may not be accepted universally, and even though most state-of-the-art document authoring and viewing tools support such fine-grained annotations, as discussed in subsection 0.2, they are rarely encouraged in peer-review workflows. To confirm the importance of this requirement, we have therefore conducted a preparatory study, which, to the best of our knowledge,

---

[13] Raw data at https://git.io/vpwxZ
[14] http://schema.org/
[15] http://purl.org/spar/deo/
[16] http://ontoware.org/swrc/
[17] https://goo.gl/9Bm2yi
[18] https://github.com/OSCOSS/fiduswriter-htmlrdfa

*4.1 "Overview"*
*Table 2 the first 3 columns reproduce table 1, which wastes space. Think of a better way of arranging this information.*
**Question**: Was the reviewer correct? – **Answer**: Yes ☐ No ☐ I am not sure ☐

**Figure 3.** A reviewer's comment and a question about it has not been conducted before: comparing the usefulness of reviewers' comments embedded into articles in a fine-grained way to the traditional practice of publishing reviews as one long block of text.

Study participants:

Of the researchers attending TPDL 2017, the Conference on Theory and Practice of Digital Libraries, we invited 6 persons to participate in this survey and divided them into two groups.

Study method:

We first selected 3 reviewed articles from the Semantic Web Journal[19], an open peer review journal established in the Web community, which had both reviews with fine-grained comments (implemented as PDF annotations) and traditional all-in-one review texts. We then chose from the reviews 10 comments overall that were referring to a section of the article and that had been given both by a reviewer making fine-grained annotations and a reviewer using traditional techniques. We designed questions about these comments, such that answering them would require the participants to read both the reviewer's comment and the part of the article addressed by the comment. We presented the articles and the reviewers' comments to the test subjects in two forms. In the first form, ranges of the article text were highlighted in the place that a reviewer had commented on, and the comments were displayed next to these highlights. (Where the reviewers had not themselves done so by using PDF annotations, *we* annotated the article accordingly.) The second form was the traditional display of the article without any highlights, and the reviews on a separate page. We assigned to each group of study participants one of the two formats of reviewed articles. We asked them to answer the questionnaire and we recorded the time it took them to answer the 10 questions. As an example, one of these questions is shown in Figure 3 The list of all selected articles and their reviews along with the questionnaire is available online.[20]

Evaluation Result:

Comparing the time consumed by the two groups of study participants shows that reviews that are connected to the parts of the article they address in a fine-grained way are faster to consume than separated reviews. The group that was provided fine-grained review comments finished their task in 9 minutes on average, whereas the group that was provided separated reviews finished their task in 20 minutes on average. Also, the participants found it more difficult to answer questions when the reviews were not attached to the affected parts of articles. These observations confirm the relevance of requiring that our proposed system should support the publication of fine-grained review comments attached to parts of articles.

## 5. Conclusion

Aiming at opening scholarly communication and specifically at opening the peer review process, we introduced a framework that allows opening a peer review workflow that is supported by the FidusWriter academic word processor and the OJS review management system. Once an article has been "finished", our method exports it with the reviewers' comments and feedback attached to it, to the dokieli HTML+RDFa format. Technically, this process is as straightforward as implementing a variant of one of the existing Fidus Writer export filters to HTML, DOCX or Open Document, but the current export filters of FW do not export comments. Thus, we developed an exporter that adds this feature.

---

[19] http://www.semantic-web-journal.net
[20] https://github.com/OSCOSS/AR-Annotator/tree/master/Evaluation

Furthermore, our method produces dokieli articles from articles that are made using Fidus writer and enriches them with standard ontologies. This promotes semantically linked open data and makes an article machine readable.

As future work, we plan to extend the current semi-automated analysis setup to a system that supports automated review analysis. Such a system, besides extracting RDFa data from a paper, directly performs queries and shows the results in the article.

Secondly, we plan to apply AR-Annotator on a corpus of articles to extract RDF datasets form articles at a large scale and integrate the extracted information with the Scholarly Communication Metadata Knowledge Graph we generated previously [5].

*Acknowledgments:* This work is developed in the context of OSCOSS project and has been partially funded by DFG under grant agreement AU 340/9-1.

1. Ross-Hellauer, T. What is open peer review? A systematic review [version 2; referees: 4 approved]. *F1000Research* **2017**, *6*. doi:10.12688/f1000research.11369.2.
2. Janowicz, K.; Hitzler, P. Open and transparent: the review process of the Semantic Web journal. *Learned Publishing* **2012**, *25*, 48–55.
3. Walker, R.; Da Silva, P.R. Emerging trends in peer review—a survey. *Frontiers in neuroscience* **2015**, *9*.
4. Walsh, E.; Rooney, M.; Appleby, L.; Wilkinson, G. Open peer review: a randomised controlled trial. *The British Journal of Psychiatry* **2000**, *176*, 47–51.
5. Sadeghi, A.; Lange, C.; Vidal, M.E.; Auer, S. Integration of Scholarly Communication Metadata Using Knowledge Graphs. Research and Advanced Technology for Digital Libraries; Kamps, J.; Tsakonas, G.; Manolopoulos, Y.; Iliadis, L.; Karydis, I., Eds.; Springer International Publishing: Cham, 2017; pp. 328–341.
6. Sadeghi, A.; Ansari, M.J.; Wilm, J.; Lange, C. A Survey of User Expectations and Tool Limitations in Collaborative Scientific Authoring and Reviewing. *arXiv preprint arXiv:1804.07708* **2018**.
7. Kriegeskorte, N. Open evaluation: a vision for entirely transparent post-publication peer review and rating for science. *Frontiers in computational neuroscience* **2012**, *6*.
8. Sadeghi, A.; Wilm, J.; Mayr, P.; Lange, C. Opening Scholarly Communication in Social Sciences by Connecting Collaborative Authoring to Peer Review. *Inf. Wiss. & Praxis* **2017**, *68*, 163. doi:10.1515/iwp-2017-0030.
9. Capadisli, S.; Guy, A.; Verborgh, R.; Lange, C.; Auer, S.; Berners-Lee, T. Decentralised Authoring, Annotations and Notifications for a Read-Write-Web with dokieli. Web Engineering; Cabot, J.; de Virgilio, R.; Torlone, R., Eds. Springer Verlag, 2017, number 10360 in Lecture Notes in Computer Science.
10. Luczak-Rösch, M.; Heese, R. Linked Data Authoring for Non-Experts. LDOW, 2009.
11. Khalili, A.; Auer, S.; Hladky, D. The RDFa content editor-from WYSIWYG to WYSIWYM. Computer Software and Applications Conference (COMPSAC), 2012 IEEE 36th Annual. IEEE, 2012, pp. 531–540.
12. Ciccarese, P.; Ocana, M.; Clark, T. Open semantic annotation of scientific publications using DOMEO. *Journal of biomedical semantics* **2012**, *3*, S1.
13. Comeau, D.C.; Islamaj Doğan, R.; Ciccarese, P.; Cohen, K.B.; Krallinger, M.; Leitner, F.; Lu, Z.; Peng, Y.; Rinaldi, F.; Torii, M.; others. BioC: a minimalist approach to interoperability for biomedical text processing. *Database* **2013**, *2013*, bat064.
14. Suhrbier, L.; Kusber, W.H.; Tschöpe, O.; Güntsch, A.; Berendsohn, W.G. AnnoSys—implementation of a generic annotation system for schema-based data using the example of biodiversity collection data. *Database* **2017**, *2017*, bax018.
15. Nuzzolese, A.G.; Peroni, S.; Recupero, D.R. ACM: article content miner for assessing the quality of scientific output. Semantic Web Evaluation Challenge. Springer, 2016, pp. 281–292.